\documentclass[twocolumn,preprintnumbers,amsmath,amssymb]{revtex4}

\usepackage{epsfig}
\usepackage{epstopdf}
\usepackage{graphicx}
\usepackage{dcolumn}
\usepackage{bm}
\usepackage{subfigure}

\begin{document}

\title{Testing Local Anisotropy Using the Method of Smoothed Residuals I - Methodology}

\author{${}^{a}$Stephen Appleby, ${}^{a,b}$Arman Shafieloo}

\affiliation{${}^{a}$Asia Pacific Center for Theoretical Physics, Pohang, Gyeonbuk 790-784, Korea \\
${}^{b}$Department of Physics, POSTECH, Pohang, Gyeongbuk 790-784, Korea}

\date{\today}

\begin{abstract}
We discuss some details regarding the method of smoothed residuals, which has recently been used to search for anisotropic signals in low-redshift distance measurements (Supernovae). In this short note we focus on some details regarding the implementation of the method, particularly the issue of effectively detecting signals in data that are inhomogeneously distributed on the sky. Using simulated data, we argue that the original method proposed in Colin et al. \cite{Colin:2010ds} will not detect spurious signals due to incomplete sky coverage, and that introducing additional Gaussian weighting to the statistic as in \cite{Feindt:2013pma} can hinder its ability to detect a signal. Issues related to the width of the Gaussian smoothing are also discussed.

\end{abstract}

\maketitle % --------------------------------------------------------------------

\section{\label{sec:i}Introduction} 

The possible existence of anisotropic signals in low redshift cosmological data sets, most notably type Ia Supernova (SN) data \cite{Riess:1998cb,Perlmutter:1998np,Amanullah:2010vv}, has recently gained some interest in the literature \cite{Kolatt:2000yg,Bonvin:2006en,Gordon:2007zw,Schwarz:2007wf,Gupta:2007pb}. Increasingly precise measurements and improved understanding of systematics has made detecting local bulk velocities of order $V_{\rm bulk} \sim 300{\rm km s^{-1}}$ up to redshifts $z \sim 0.1$ possible (albeit with weak significance). A number of methods have been proposed to detect anisotropic signals of both cosmological \cite{Appleby:2012as,Koivisto:2008ig,Koivisto:2010dr,Blomqvist:2008ud,Cooray:2008qn,Gupta:2010jp,Cooke:2009ws,Antoniou:2010gw} and local origin \cite{Colin:2010ds,Feindt:2013pma}. Cosmological anisotropy is severely restricted by the uniformity of the distance to last scattering on the sky \cite{Campanelli:2006vb,Campanelli:2007qn,Appleby:2009za}, however a local bulk flow would introduce an anisotropic signal at low redshifts exclusively. 

An open question regards the scale of any potential bulk velocity - whether there is coherent motion amongst the data or local flows in the vicinity of large overdensities. In the former case, the effect of the bulk velocity on observables can be calculated theoretically, and the resulting model directly compared to the data. However local bulk flows due to inhomogeneous matter distributions are difficult to model. Model independent tests of the isotropic hypothesis are therefore particularly useful, since one can obtain information regarding the magnitude of $V_{\rm bulk}$, its direction on the sky and the scale over which it acts. 

In this work we would like to discuss some technical issues related to one particular approach to studying anisotropy - the use of the so called `method of smoothed residuals'. We begin by reviewing the original method, outlined in \cite{Colin:2010ds}. We also review a slightly modified form of the method, proposed in a recent paper \cite{Feindt:2013pma}. Whilst the two approaches are similar, there are important differences that we would like to highlight. We focus on the use of both methods when the data is not homogeneously distributed on the sky, and how the original method avoids spurious detections of anisotropy in directions that are oversampled. 

The note will proceed as follows. In section \ref{sec:0} we discuss the original method of residuals, applied to a SN sample (although the arguments that we make throughout apply generally to the method). We also explain a slightly modified variant of the method, used in \cite{Feindt:2013pma}. In section \ref{sec:1}-\ref{sec:3} we discuss some details, and clarify issues related to inhomogeneous data sets, width of Gaussian smoothing and the nature of the hypothesis tests being performed. We conclude in section \ref{sec:IV}. In the appendix we provide a brief order of magnitude calculation of the signal and noise expected in current surveys.

\section{\label{sec:0}Method of Smoothed Residuals}

We begin with a review of the original method proposed in \cite{Colin:2010ds}. We apply it to a Supernova data sample, however it should be considered as a general method for detecting anisotropic signals in data on the sky. The analysis involves the following steps:

First we calculate a global best fit value assuming a model (here flat $\Lambda$CDM with parameters $H_{0}$ and $\Omega_{\rm m0}$)\footnote{Our results will be only weakly dependent upon the underlying model chosen, since we are exclusively using a low redshift data set.}, by minimizing the $\chi^{2}$ statistic using the distance moduli $\mu_{\rm i}$ for the $i=1,N$ data points. Using these parameters, we construct the error-normalised difference of the data from the model \cite{Perivolaropoulos:2008yc,Perivolaropoulos:2010yc,Shafieloo:2010xm},

\begin{equation} \label{eq:res1}
q_i (z_i, \theta_i, \phi_i) = \frac{\mu_i(z_i, \theta_i, \phi_i) 
- \tilde\mu(z_{\rm i})}{\sigma_i}\;, 
\end{equation}

\noindent where $\sigma_{\rm i}$ is the error on the ${\rm i}^{\rm th}$ data point. We denote the best-fit distance modulus as $\tilde\mu(z,H_{0})$. Here, $(\theta_i, \phi_i)$ are the positions of the $i$ data point on the sky
Henceforth we work with these residuals, $q_i (z_i, \theta_i,
\phi_i)$ and consider their angular distribution on the sky.

Next, we define a measure $Q (\theta, \phi)$ on the surface of a
sphere of unit radius using these residuals:
\begin{equation}
Q (\theta,\phi) = \sum_{i=1}^N q_i (z_i, \theta_i, \phi_i)
W (\theta, \phi, \theta_i, \phi_i)\;, 
\end{equation} 
where $N$ is the number of SNe\,Ia data points and $W (\theta, \phi,
\theta_i,\phi_i)$ is a window function that represents a two dimensional
smoothing over the surface of a sphere. We define the window function using a
Gaussian distribution:

\begin{equation}
W (\theta, \phi, \theta_i, \phi_i) = \frac{1}{\sqrt{2\pi}\delta} 
\exp\left[-\frac{L(\theta, \phi, \theta_i, \phi_i)^2}{2\delta^2}\right]\;,
\end{equation} 

\noindent where $\delta$ is the width of smoothing and $L (\theta, \phi,
\theta_i, \phi_i)$ is the distance on the surface of a sphere of unit
radius between two points with spherical coordinates $(\theta, \phi)$
and $(\theta_i, \phi_i)$. The Gaussian window function is not unique; for example one could instead opt for a top hat 
profile \cite{Kalus:2012zu}. However we found no significant change to our results when repeating our analysis 
using this different function.

Any anisotropy in the data will
translate to $Q (\theta, \phi)$ significantly deviating from zero. Note that the expectation value of $Q(\theta,\phi)$ is non-zero even when we fit the `correct' model to the data, but its value will have no statistical significance in this case.

Finally we adopt a value for $\delta$, calculate $Q (\theta, \phi)$ over the
whole surface of the sphere and find the minimum and maximum of this
quantity, $Q_{\rm min} = Q (\theta_\mathrm{min}, \phi_\mathrm{min})$ and $Q_{\rm max} = Q
(\theta_\mathrm{max}, \phi_\mathrm{max})$. Our statistical measure of
anisotropy is based on the difference
\begin{equation} 
\Delta Q = Q (\theta_\mathrm{max}, \phi_\mathrm{max}) 
- Q (\theta_\mathrm{min}, \phi_\mathrm{min}) ,
\label{DeltaQ}
\end{equation}
i.e. a large value of $\Delta Q_\mathrm{data}$ relative to $\Delta Q$ obtained from fiducial mock data sets implies significant
anisotropy. In this work, we will use $\Delta Q$ when calculating the significance of any bulk velocity detection. However when we plot the direction of any detection of anisotropy on the sky we always use the direction associated with $Q_{\rm max}$, since this should be aligned with the bulk velocity in our simulations. 

The function $Q(\theta,\phi)$ constitutes a map of the data over the whole sky, and as such contains significantly more information that one can extract just using $\Delta Q$. One could contruct any number of different measures to test various hypotheses, however in this work we concentrate exclusively on $\Delta Q$ by virtue of its simplicity and its clear ability to discriminate anisotropic signals \cite{Colin:2010ds}.

To calculate the significance of the $\Delta Q$ value obtained from the data, one must construct a distribution of $\Delta Q$ values obtained by simulating $N_{\rm real}$ data sets based upon the null hypothesis being tested. In the original method, $N_{\rm real}=10^{3}$ realisations were simulated, with the same positions, redshifts and errors as the data. The distance modulus of each point $\mu_{i}$ was taken to be the data best fit value $\tilde{\mu}(z,H_{0})$ plus a contribution drawn from a Gaussian of width $\sigma_{\rm i}$. The $\Delta Q$ value obtained from the data is compared to the resulting distribution of $\Delta Q$ from the $N_{\rm real}$ simulations, to test the significance of the result. 

There are two important points that must be stressed. The first is that if there is no bulk velocity or anisotropy in the data and systematics are under control, then one can expect that every residual will be drawn from a symmetric distribution, and this is true regardless of the positions of the data points. This is the null hypothesis that the original method was designed to test. The second is that the simulated data has the same distribution on the sky as the actual data. 

A variant of the method was discussed in \cite{Feindt:2013pma}. There, the approach followed the original to a large degree. The best fit cosmology is calculated by minimizing $\chi^{2}$, and the data residuals are calculated by subtracting $\tilde{\mu}$ from the data values. A modified quantity was introduced, which we call $\bar{Q}$, defined as 

\begin{equation} \label{eq:100}
\tilde{Q} (\theta,\phi) = { \sum_{i=1}^N q_i (z_i, \theta_i, \phi_i)
W (\theta, \phi, \theta_i, \phi_i) \over \sum_{j=1}^N
W (\theta, \phi, \theta_j, \phi_j) } \;, 
\end{equation} 

\noindent that is, $Q$ obtained from the original method weighted by the sum of the window function at that point. $\Delta \tilde{Q} = \tilde{Q}_{\rm max}-\tilde{Q}_{\rm min}$ is similarly defined. 

To find the significance of the magnitude of $\Delta \tilde{Q}$, one must construct a distribution based upon the hypothesis being tested. In \cite{Feindt:2013pma}, the authors create a distribution of $\Delta \tilde{Q}$ by taking the $N$ residuals and redistributing them randomly amongst the data point positions $N_{\rm real}=5000$ times, each time calculating $\Delta \tilde{Q}$.

We call these two methods A and B respectively. In the following section we discuss how the two methods differ. Before continuing we stress that this note is not a criticism of the main results of \cite{Feindt:2013pma} - here we make no claim as to the direction or magnitude of the local bulk flow in actual supernova data. A thorough analysis using the method of smoothed residuals applied to various available data sets will follow this work. Here we are simply discussing the method and its application.

\section{Discussion}

Methods A and B represent slightly different implementations of the method of smoothed residuals, and in this section we discuss some of the differences. In what follows we will support our arguments by calculating distributions of $\Delta Q$ for simulated data sets. For the simulations we use a background $\Lambda$CDM cosmology with $h_{0,d}=0.7$, $\Omega_{\rm m0, d}=0.27$. In each simulation we add statistical noise to the distance modulus data, $\delta \mu_{\rm G}$, drawn from a Gaussian of width $\sigma_{\mu} = 0.13$. For any given distribution of points we create two sets of $N_{\rm real} \sim 10^{3}$ mock data, one with a non-zero global bulk velocity $\tilde{v}$ of magnitude $V_{\rm bulk}=250 {\rm km s^{-1}}$ and a fiducial $\Lambda$CDM case with $V_{\rm bulk} = 0$. The bulk velocities direction is chosen arbitrarily as $(l,b)=(30,120)$. Unless stated otherwise, we use $N=300$ data points, which we distribute between $z=(0.015,0.1)$ with a flat prior.

\subsection{\label{sec:1}Spurious Signals due to Inhomogeneous Data}

The main point of this note is to refute the suggestion that the original method A will detect spurious anisotropic signals if the data is inhomogeneously distributed. We do not expect this to be the case, due to the fact that one compares the observed value of $\Delta Q$ from the data with the distribution of $\Delta Q$ values obtained from mock data with {\it the same data positions on the sky} for each realisation. Therefore any effect on $\Delta Q$ due to the distribution of points will be observed in both the realisations and data. In this respect, there is no need to weight the original $\Delta Q$ function as in method B - in fact in certain data distributions weighting the $\Delta Q$ function might reduce the significance of a signal relative to the noise. 

The statistical significance of any anisotropic signal obtained with method A is based upon the $\Lambda$CDM, $V_{\rm bulk}=0$ realisations drawn from the same point distribution on the sphere. As such the distribution does not play a significant role - the variance of the statistical noise from the realisations will be a function of the number of data points in a given region. 

We discuss a small number of sample cases. The simplest - an isotropic distribution of $N$ data on the sky - is optimal for detecting a bulk flow, both in terms of the significance of the magnitude of $\Delta Q$ and in correctly locating the direction $(\theta_{\rm max},\phi_{\rm max})$. This can be seen by examining the residuals. If we make the simplifying assumptions that the bulk velocity has a small perturbative effect on the Hubble parameter and the data is located at small $z \ll 1$ \footnote{Neither of these assumptions are made when we perform our numerical calculations - they are made purely to simplify the analytic results and relaxing them does not alter our conclusions}, then we can write the sum of residuals as 

\begin{equation} \sum_{\rm i=1}^{\rm N} q_{\rm i}(z_{\rm i},\theta_{\rm i},\phi_{\rm i}) \simeq \sum_{\rm i=1}^{N} \left( \delta \mu_{\rm G, i} + A_{0} {\tilde{v}.\hat{n}_{\rm i} \over cz_{\rm i}}\right) + A_{0} N\delta H \end{equation}

\noindent where $\delta H = H_{\rm 0, th}/H_{\rm 0 d}-1$, with $H_{\rm 0, th}$ being the best fit $H_{0}$ value obtained by $\chi^{2}$ minimization, and $H_{0,d}$ is the `true' value used in the simulations. Here $A_{0}$ is an unimportant constant of order unity. If we make the further simplifying assumption that the sum of residuals is zero (this is not exactly the case) then one can estimate $\delta H$ as 

\begin{equation} \delta H \sim -{1 \over N} \sum_{\rm i=1}^{N} \left( { \delta \mu_{\rm G, i} \over A_{0}} + {\tilde{v}.\hat{n}_{\rm i} \over cz_{\rm i}}\right) \end{equation}

\noindent The sum of the Gaussian noise can be considered as a single variable drawn from a Gaussian of zero mean, and for an isotropic distribution the sum $\sum_{\rm i}\tilde{v}.\hat{n}_{\rm i}/z_{\rm i}$ will have expectation value zero regardless of the magnitude $V_{\rm bulk}$. This data distribution will yield a set of residuals that is symmetrically distributed around the `true' underlying cosmological model. Let us compare this case to one which might naively be expected to yield a stronger signal in $\Delta Q$. We now take $N=200$ points isotropically distributed on the sky and $N_{\rm patch} = 100$ points densely located in a region $l=(20,40)$, $b = (110,130)$ (that is, in the direction of the simulated bulk velocity). In such a case, $\delta H$ will acquire a non-zero expectation value of order 

\begin{equation} \langle \delta H \rangle \sim -{N_{\rm patch} \over N} { V_{\rm bulk} \over c \bar{z}} \end{equation}

\noindent where $\bar{z}$ is the average redshift of the data points in the patch (this is an order of magnitude approximation only). We see that increasing the number of points in the direction of the anisotropy has the effect of shifting the best fit cosmological model away from the underlying one. Despite this, increasing the number of points in the specific direction of the bulk velocity will generically have the effect of increasing the signal, despite the penalty in shifting the best fit function. While it is intuitively obvious that increasing the number of data points in the direction of the bulk flow will increase the significance of the detection, it is also clear that the signal per data point will decrease for an inhomogeneous distribution. Hence an isotropic distribution is optimal if we make no prior assumptions regarding the direction of the bulk velocity. 

To confirm our reasoning, we resort to simulations. We create six sets of $N_{\rm real} =10^{3}$ mock data containing $N=300$ points. In two simulations we distribute the three hundred data isotropically on the sky, two sets are created with a subset of $N_{\rm patch}=100$ points distributed equally in two patches $\theta = (20,40)$, $\phi=(110,130)$ and $\theta = (-20,-40)$, $\phi=(290,310)$ (that is, in the direction of the bulk velocity and its dipole) and in the final two we have $N_{\rm patch}=100$ points distributed in a patch in an arbitrary direction, taken to be $\theta = (-50,-70)$, $\phi=(30,50)$. For each of the three distributions, one set of simulations has $V_{\rm bulk}=250{\rm km s^{-1}}$ and the other $V_{\rm bulk} = 0$. For each realisation we calculate $\Delta Q$. If there is a bulk velocity, then $\Delta Q$ should be significantly larger than its value when $V_{\rm bulk}=0$. As a null test of both methods, we performed simulations of an isotropic cosmological model with the third distribution described above. We found no statistically significant spurious signals, as expected.

In table \ref{tab:1} we exhibit the number of $V_{\rm bulk}=250{\rm km s^{-1}}$ simulations that have a $\Delta Q$ that is larger than $99\%$ of the $V_{\rm bulk}=0$ realisations $\Delta Q$ values. In other words, these are the $V_{\rm bulk} = 250 {\rm km s^{-1}}$ simulations which can correctly exclude the zero bulk velocity hypothesis at $99\%$ confidence. We compare the number of successful discriminations between an isotropically distributed data set and one containing a patch of data at an arbitrary location. The isotropic data distribution provides a more reliable test of anisotropy, in the sense that more isotropic realisations are excluded than when we have an inhomogeneous distribution in an arbitrary direction. 

\begin{figure*}
\centering
\mbox{\resizebox{0.85\textwidth}{!}{\includegraphics[angle=270]{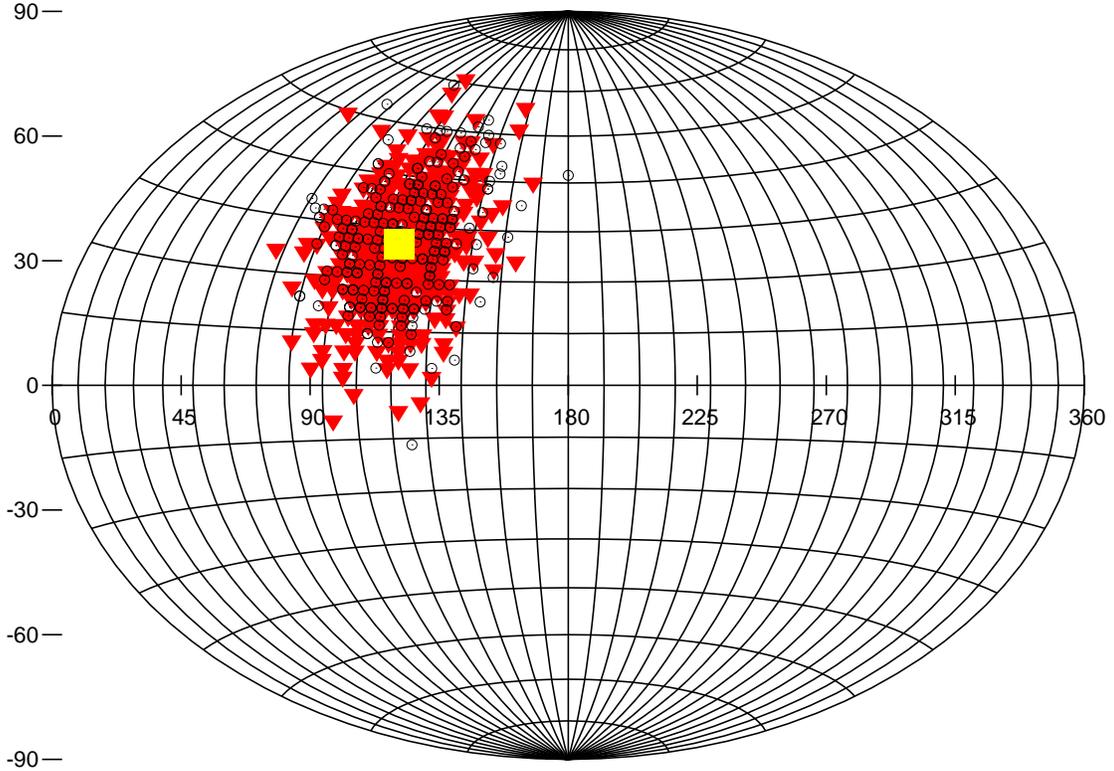}}} 
\caption{The realisations of simulated $V_{\rm bulk}=250 {\rm km s^{-1}}$ mock data that exclude the isotropic $V_{\rm bulk}=0$ simulations to $99\%$ confidence. Both sets of data are for $N=200$ points isotropically distributed on the sky, and $N_{\rm patch} = 100$ points equally populated in the vicinity of the bulk velocity $(\theta,\phi)=(30,120)$ and its dipole $(\theta,\phi)=(-30,300)$. The red triangles ($N_{\rm exc}=838$) denote the $Q_{\rm max}$ statistic used in method A, and the black dots ($N_{\rm exc}=189$) the modified $\bar{Q}_{\rm max}$ statistic of method B. The yellow square denotes the actual direction of the bulk velocity. The discriminatory nature of the test is reduced when we divide out the window function for reasons discussed in the text. 
}
\label{fig:1} 
\end{figure*}

For the case where we have a patch of data in the same direction as the bulk velocity, one can argue that introducing the weighting in method B actually scatters the direction of $Q_{\rm max}$, and hence reduces the utility of the method. For method A the signal will clearly be the largest in the direction of the bulk velocity - the signal grows roughly linearly with the number of points in a given direction and this is true regardless of $\delta H$ corrections to the signal. Since method B uses $\bar{Q}$ defined in ($\ref{eq:100}$), it divides by the sum of the window function at each point. Since this weighting will be largest in the location of the overdense patch (by virtue of it having the largest number of points), {\it at best} method B can find the location of the bulk velocity as well as method A. However in actuality the method will introduce additional scatter in the detection. This is exhibited in fig.\ref{fig:1} - the black points correspond to the $\Delta Q$ values that rule out $V_{\rm bulk}=0$ to $99\%$ confidence for method B, and the red points method A. There are considerably fewer black points than red, indicating that the actual significance of the signal is reduced by dividing out the window function. Note that both methods will pick out the correct direction here, since we are using a unique inhomogeneous data set where the overdense data region is aligned with $\tilde{v}$. 

Some additional comments. Although method B is inferior to method A when the inhomogeneous data patch is aligned with the bulk velocity, this is not a generic statement. One can create scenario's where the converse is true. This highlights two issues. One - the question of accurate directional detection is complicated considerably by a non-trivial distribution of data. There is no single variant of the $Q$ function that will optimally detect the bulk velocity direction for an arbitrary distribution. Given a distribution an optimal weighting could be constructed, however this would require extensive simulations. In this respect, one should consider the smoothed residual method as it was originally intended; as a null test of the isotropic hypothesis. Two - method A will not detect spurious bulk velocity signals due to inhomogeneities in the data distribution, since we are comparing data and realisations with the same positions on the sky. It is a reasonable to state that the sensitivity of the method will be reduced for an inhomogeneous distribution relative to homogeneous data. However this problem is not necessarily ameliorated by dividing out the window function at each point.

%%%%%%%%%%%%%%%%%%%%%%% 
\begin{table}[!htb]
\begin{tabular}{l|ccc} 
Distribution \ & \ $N_{\rm exc}$ (method A) \ & $N_{\rm exc}$ (method B) & \ \\
\hline 
Isotropic & 404 & 406 & \\ 
Anisotropic I & 239 & 212 & \\ 
Anisotropic II & 838 & 189 & \\ 
\end{tabular}
\caption{Number of $V_{\rm bulk}=250{\rm km s^{-1}}$ realisations that excludes the $V_{\rm bulk}=0$ hypothesis to $99\%$ confidence. We perform $N=10^{3}$ realisations in total for each case. Isotropic is the case where we have $N=300$ data points isotropically distributed on the sky, Anisotropic I is for $N=200$ points isotropically distributed and $N_{\rm patch}=100$ points densely packed in the vicinity $(\theta,\phi) = (-60,40)$ (arbitrarily chosen), and Anisotropic II is $N=200$ points isotropically distributed and $N_{\rm patch}=100$ points equally distributed in two patches in the vicinity of the bulk velocity $(\theta,\phi) = (30,120)$ and its dipole $(\theta,\phi) = (-30,300)$. One can see that the two methods perform comparably in the first two cases, and method A is clearly more discriminatory when we have many data points in the direction of the signal. This is due to the fact that the signal is cumulative in the original method.
}
\label{tab:1}
\end{table}

\subsection{\label{sec:2}Width of Gaussian Smoothing}

The second important point that we wish to make is that the window function contains a free parameter, $\delta$, which is the scale of the Gaussian smoothing. The sensitivity of the method to different anisotropic signals will depend on $\delta$. For example, if we are searching for a global bulk velocity in the SN data, then $\delta = \pi/2$ will be optimal in detecting the signal. However, if we are searching for local anisotropies, $\delta =\pi/2$ will not necessarily be efficient in detecting them. Ideally, one should choose the value of $\delta$ to be the size of the angle on the sky over which the anisotropic signal is expected to be significant. If we have no prior knowledge of the anisotropic signal, one should perform an analysis for multiple $\delta$ values and compare the significance of the resulting $\Delta Q$ to mock data in each case. Finding the value of $\delta$ that yields the maximal significance in $\Delta Q$ will give us additional information regarding the distribution of bulk velocities in the data (that is, whether the bulk velocity is global or local, and if local on what typical scale it is present). This information is important, as it would be very difficult to construct a theoretical model of local bulk velocities that we can use to directly confront with the data. 

\begin{figure*}
\centering
\mbox{\resizebox{0.45\textwidth}{!}{\includegraphics[angle=270]{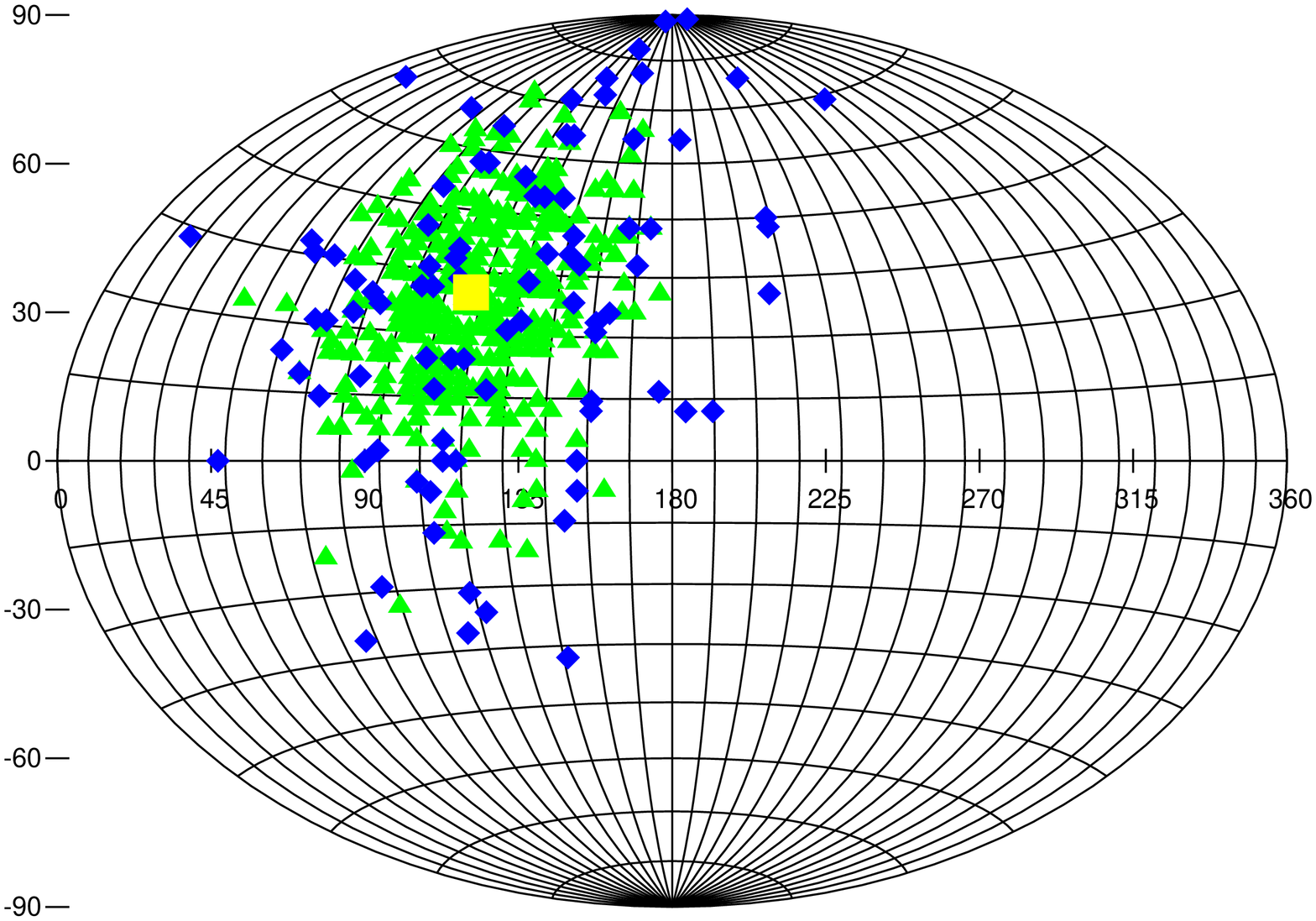}}}
\mbox{\resizebox{0.45\textwidth}{!}{\includegraphics[angle=270]{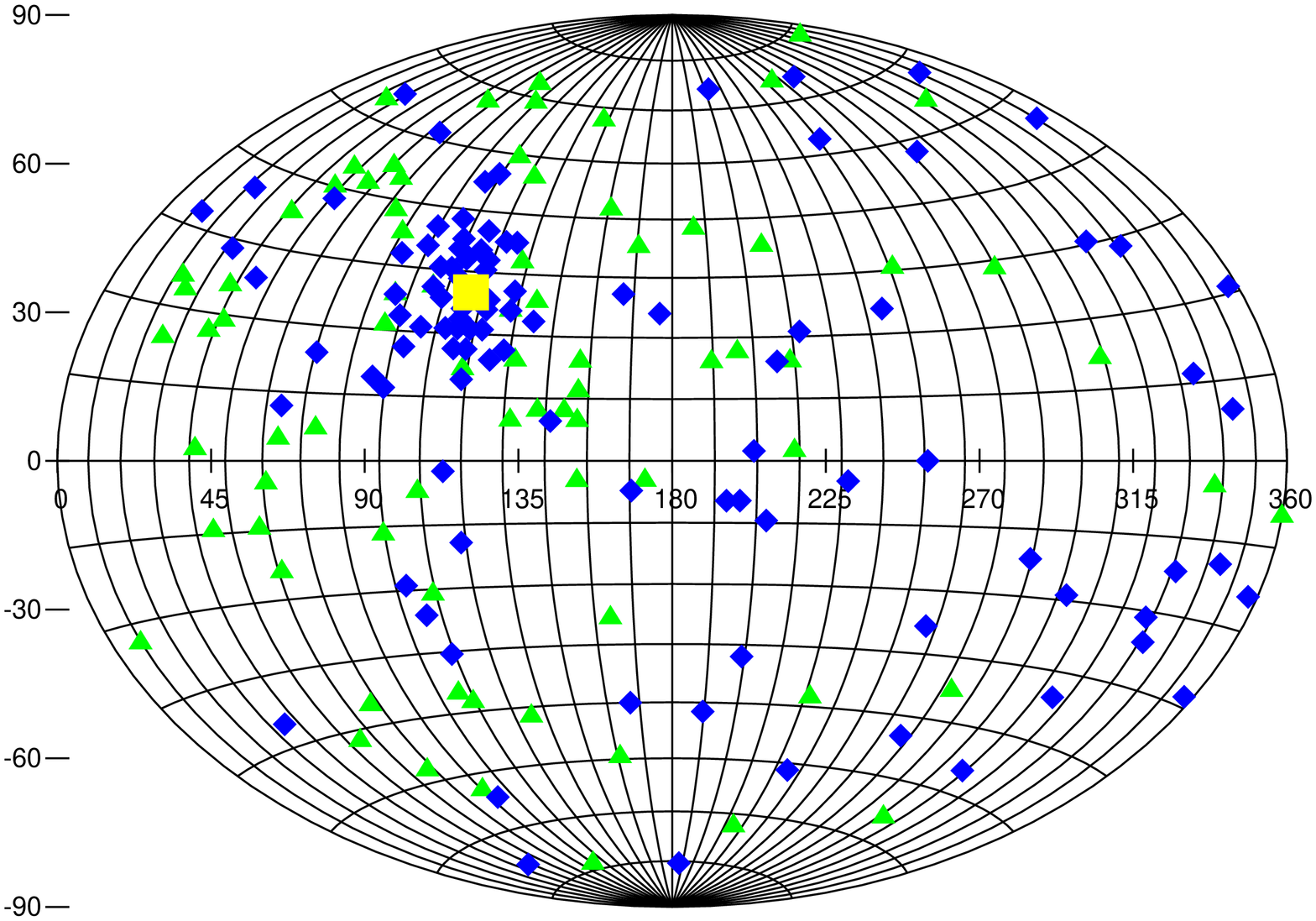}}} 
\caption{[Left Panel] Realisations that discriminate between a global bulk velocity $V_{\rm bulk}=250{\rm km s^{-1}}$ and $V_{\rm bulk}=0$ at $99\%$ confidence. The green triangles are for a window function with $\delta =\pi/2$ and the blue diamonds are $\delta =\pi/9$. There are fewer points as we decrease the smoothing scale, reflecting the fact that for a global velocity, the bulk flow signal is cumulative over an entire hemisphere. [Right Panel] Realisations that discriminate between a local bulk velocity that only affects points in the range $\theta=(20,40)$, $\phi=(100,140)$. We are now using $N=500$ data points and $V_{\rm bulk}=600{\rm km s^{-1}}$. The increased scatter is due to the fact that we are now looking at the $95\%$ confidence limit. The green triangles represent the Gaussian smoothing $\delta=\pi/9$ and the blue $\delta=\pi/2$. Now the converse is true - the local bulk velocity is preferentially selected by the narrow smoothing scale. Enlarging $\delta$ now has the effect of increasing the noise with no gain in signal. Note however that it is much more difficult to detect a local bulk velocity than global - there are considerably fewer successful discriminations compared to the global bulk velocity case. 
}
\label{fig:2} 
\end{figure*}

In the left panel of fig.\ref{fig:2} we exhibit the realisations of $V_{\rm bulk}=250{\rm km s^{-1}}$ data that preclude the $V_{\rm bulk}=0$ realisations to $\sim 99\%$ confidence for an $N=300$ isotropic data distribution and a global velocity. The green points denote a smoothing scale of $\delta = \pi/2$ and the blue diamonds $\delta = \pi/9$. We use the original method A exclusively in this subsection. There are clearly fewer successful discriminations in the smaller $\delta$ smoothing, reflecting the fact that on smaller scales we pick up a smaller cumulative signal. For a global bulk velocity the signal will be coherent over an entire hemisphere. 

Contrast this to the right panel of fig.\ref{fig:2}, where we now distribute $N=500$ points isotropically on the sky, and provide a local bulk velocity $V_{\rm bulk}=600 {\rm km s^{-1}}$ to the points that fall in the range $\theta = (20,40)$, $\phi=(100,140)$. Note that here we are exaggerating both the magnitude of the bulk velocity and the number of data points to detect the signal - a local bulk velocity is considerably harder to detect than a global bulk flow. We repeat our analysis with $\delta =\pi/2$ (green points) and $\delta = \pi/9$ (blue diamonds). We now plot the realisations that exclude the isotropic hypothesis to $95\%$ significance, hence the increased scatter. One can see that there are fewer successful detections of the velocity compared to the global bulk velocity (left panel). This is a reflection of the reduced signal associated with a local flow. However, it is clear that the value of $\delta$ that encompasses the local patch, $\delta =\pi/9$, performs better than the $\delta = \pi/2$ scale. This is due to the fact that when we use $\delta = \pi/2$, we increase the number of points in a patch (and hence the noise) but not the signal, which is restricted to a small region of the sky.

\subsection{\label{sec:3}Hypothesis Testing}

We make an additional comment on the actual statistical tests being used in \cite{Colin:2010ds} and \cite{Feindt:2013pma}. In both methods, the quantity $\Delta Q$ (or $\Delta \bar{Q}$) is obtained from the data in the same way. However the distributions that they are then tested against are different, and hence the hypotheses being tested in the two cases are distinct. In method A, the realisations are drawn from an isotropic $\Lambda$CDM Universe with Gaussian statistical errors and $V_{\rm bulk}=0$. Method B takes the residuals obtained from the data and resamples them in random data point directions (keeping the data point locations fixed). One can think of this roughly as drawing residuals from their empirical distribution function constructed from the data. The method does not assume that these are Gaussian distributed around the best fit model, and they will generically not be. If there is an underlying bulk velocity and the data is not isotropically distributed on the sky, then one can say that while the residuals might sum to the value that minimizes $\chi^{2}$, the distribution will not necessarily be symmetric around this value. 

Therefore one can state the hypothesis being tested in method B as `The value of $\Delta Q$ obtained from the data is consistent with the data residuals being randomly distributed amongst the data point positions'. This is not the hypothesis of method A, which can be stated as - `The value of $\Delta Q$ obtained from the data is consistent with Gaussian statistical fluctuations in an isotropic Universe'. The two are mathematically equivalent only when the bulk velocity is zero.

Both hypotheses are tests of isotropy, but they are distinct. In effect, there are actually two tests of isotropy involved here - whether the residuals are distributed symmetrically around the model that minimizes $\chi^{2}$ and whether the positions of the residuals on the sky are uncorrelated (the $\Delta Q$ value essentially tests the directional relationship between the residuals). If either of these assumptions are violated, method A will pick up a signal in $\Delta Q$, whereas method B can only distinguish potential correlations in the residual's positions. In practice the distinction might be small, however upcoming surveys will yield unprecedented constraints on the distance measurements, making the distribution of the residuals an additional discriminant when searching for a bulk velocity.

\section{\label{sec:IV}Conclusions}

To conclude, in this short note we have discussed a number of issues relating to the method of smoothed residuals. The main point is to state that the original method \cite{Colin:2010ds} does not require modification to avoid spurious detections due to inhomogeneous sampling. Any sampling effect is taken into account when we compare the $\Delta Q$ value to mock data sets with the same distribution on the sky. If we have a patch of the sky containing more data, then the mean of the residuals in this patch should still be zero if there is no bulk velocity. The statistical noise has a higher probability of being large in this region, but this is not relevant as the same elevated noise will be present in the $V_{\rm bulk}=0$ simulations against which we test the significance. If there is a non-zero bulk velocity, then the additional points might shift the detected direction of the bulk flow away from the true direction. However the method is intended as a null hypothesis test - we are primarily interested in determining the existence of a bulk velocity. In this respect introducing an additional weighting actually hinders the method, as the signal would no longer be cumulative.

It is important to stress that the ability of either method A or B to detect a signal in the data depends sensitively on the distribution of data on the sky. One should test the significance of the magnitude of the $\Delta Q$ value by comparing the observed value against $V_{\rm bulk}=0$ simulations. To test the reliability of the directional detection one should additionally perform simulations with a non-zero bulk flow $V_{\rm bulk}\ne 0$. Only by performing this additional test can one deduce the effect of the distribution of data on the detected direction. The original method was intended as a null test of the isotropic hypothesis, and if we treat it as such then one need only test the magnitude of the $\Delta Q$ function. Obtaining further information requires more care and utilising a second independent method, as in \cite{Colin:2010ds,Feindt:2013pma}, is extremely prudent. 

The second important point is that the question as to whether the bulk flow is global and acts on all SN data points coherently, or only in a local patch on the sky, should be addressed by varying the parameter $\delta$. The signal to noise ratio in current data sets is marginal - the advantage of using $\delta$ is that one can maximize the signal to noise expected for a given bulk velocity. For example using $\delta =\pi/2$ will collect a signal over an entire hemisphere, making it useful for studying velocities containing a coherent signal over these scales. Varying $\delta$ will yield additional information regarding the distribution of $\tilde{v}$. 

Significant detection of a bulk velocity in low redshift data remains an open challenge, and near-future data sets will allow us to definitively answer the question as to whether the bulk flow is local or coherent over the whole sky. Model independent tests such as the method of smoothed residuals will play a central part in answering this question. However when using inhomogeneous data one must be cautious in applying the method. The original method is optimal to test the isotropic hypothesis - extracting further information such as the magnitude $V_{\rm bulk}$, direction and scale of any bulk velocity requires a much more detailed analysis. Testing the results obtained from the data by using extensive simulations, taking maximally isotropic subsamples of the data and using alternative methods as a means of independent verification is imperative.

\section*{Appendix - Signal to Noise}

We include here a brief discussion of the size of the signal expected in current Supernova surveys. Let us consider how the signal and noise varies as we vary the size of a patch on the sky. Increasing the area of the patch will increase the number of data points, increasing both the signal and the variance of the noise. Here we assume an isotropic distribution of data over the whole sky, and a global bulk velocity $\tilde{v}$ of magnitude $V_{\rm bulk}$. The residual of the $i^{\rm th}$ data point can be approximately written as 

\begin{equation} \label{eq:a1} q_{\rm i} \simeq {1 \over \sigma_{\rm i}} \left[ A_{0}{\tilde{v}.\hat{n}_{\rm i} \over cz_{\rm i}} + \delta \mu_{\rm G,i} \right] \end{equation}

\noindent where $\delta \mu_{\rm G,i}$ is drawn from a Gaussian distribution of width $\sigma_{\rm i}$. For simplicity we calculate the sum of residuals in a patch on the sky containing $N_{\rm patch}$ points, neglecting the window function $W(\theta,\phi,\theta_{\rm i},\phi_{\rm i})$

\begin{equation}\label{eq:a2} q_{\rm patch} = \sum_{i=1}^{N_{\rm patch}} q_{\rm i} \end{equation}

\noindent The function $W(\theta,\phi,\theta_{\rm i},\phi_{\rm i})$ will modify the overall $Q$ value, however here we are solely interested in the ratio of signal to noise (the first and second terms on the right hand side of ($\ref{eq:a1}$) respectively), and $W(\theta,\phi,\theta_{\rm i},\phi_{\rm i})$ weights both equally. As in the main body of this work, we take $\sigma_{\rm i}$ to be the same for all of the data points.

The sum of $N_{\rm patch}$ Gaussian contributions $\delta \mu_{\rm G,i}$ is itself a single variable drawn from a Gaussian of mean zero and variance $\sigma^{2}=N_{\rm patch}\sigma_{\rm i}^{2}$. Therefore the standard deviation of the statistical noise grows like $\sim \sqrt{N_{\rm patch}}$. The signal does not grow linearly with $N_{\rm patch}$ due to the angular dependence of $\tilde{v}.\hat{n}_{\rm i}$. 

Let us take a spherical patch on the sky that subtends an angle $2\theta$ on the sky, and is centered in the direction of the bulk velocity (so $\theta=0$ at $\tilde{v}.\hat{n} = V_{\rm bulk}$). For an isotropic distribution of points, one can find a relationship between the number of points and the area of the patch, given by $\theta$. If there are $N_{\rm tot}$ points on the whole sphere, then the variance of the noise grows like 

\begin{equation}\label{eq:f1} \sigma^{2} = {N_{\rm tot} \over 2} \left( 1-\cos[\theta] \right) \sigma_{\rm i}^{2} \end{equation}

\noindent and the signal

\begin{equation} \sum_{\rm i=1}^{N_{\rm patch}} {\tilde{v}.\hat{n} \over cz_{\rm i}} \sim {N_{\rm tot} V_{\rm bulk} \over 4 c \langle z \rangle }\left(1 - \cos^{2}[\theta]\right) \end{equation} 

\noindent where again $\theta$ denotes half the angle subtended by the spherical patch. One can see that the signal increases at a faster rate than the noise close to the direction of the bulk velocity ($\theta \sim 0$), but slower at the hemisphere ($\theta \sim \pi/2$), as one might expect. For a given survey one can use order of magnitude calculations such as this to obtain the optimal Gaussian width that we should use to detect a bulk velocity of particular magnitude $V_{\rm bulk}$ and characteristic scale. Such an approach will be utilized in a forthcoming publication.

\section*{Acknowledgements}
S.A.A and A.S wish to acknowledge support from the Korea Ministry of Education, Science and Technology, Gyeongsangbuk-Do and Pohang City for Independent Junior Research Groups at the Asia Pacific Center for Theoretical Physics. The authors would like to thank Eric Linder and Ewan Cameron for helpful discussions.

\end{document}